\documentclass[aps,prl,twocolumn,10pt,longbibliography,nobibnotes]{revtex4-2}
\usepackage{amsmath,amssymb,graphicx,mathtools,dsfont,courier}
\usepackage[T1]{fontenc}
\usepackage{xcolor}
\usepackage{ulem}

\usepackage{amsthm}

\begin{document}
\newcommand{\ketbra}[2]{|#1\rangle\!\langle#2|}
\newcommand{\braket}[2]{\langle#1|#2\rangle}
\newcommand{\tra}[1]    {{\rm tr}[ #1 ]}
\newcommand{\trr}[2]    {{\rm tr}[ #1 ]_{\overline{#2}}}
\newcommand{\titr}[1]    {\widetilde{{\rm Tr}}\left[ #1 \right]}
\newcommand{\avs}[1]    {\langle #1 \rangle}
\newcommand{\modsq}[1]    {| #1 |^2}
\newcommand{\0}    {\ket{\vec 0}}
\newcommand{\1}    {\ket{\vec 1}}
\newcommand{\dt}    {\delta\theta}
\newcommand{\I}    {\mathcal  I_{q}^{(0)}}
\newcommand{\Id}[1]    {\mathcal  I_q\left[#1\right]}
\newcommand{\IdA}    {\overline{\mathcal  I}_q^{(A)}}
\newcommand{\Ic}[1]    {\mathcal  I_{\hat A}\left[#1\right]}
\newcommand{\C}    {\hat{\mathcal C}(t)}
\newcommand{\Cd}    {\hat{\mathcal C}^\dagger(t)}
\newcommand{\re}[1]    {\texttt{Re}\left[#1\right]}
\newcommand{\im}[1]    {\texttt{Im}\left[#1\right]}

\newcommand{\ket}[1]{| #1 \rangle}
\newcommand{\bra}[1]{\langle #1 |}
\newcommand{\haarint}{\int_{U(d)}d\mu(U)}
\newcommand{\haaravg}{\mathbb{E}_{\mu(U)}}
\newcommand{\hvg}[1]{\mathbb{E}_{\mu(U)}\left[#1\right]}
\newcommand{\tr}{\mathrm{tr}}

\renewcommand{\dag}[1]{#1 ^\dagger}

\title{Volume-law protection of metrological advantage}
\author{Piotr Wysocki}
\affiliation{
  Institute for Quantum Optics and Quantum Information of the Austrian Academy of Sciences, 6020 Innsbruck, Austria
}
\affiliation{Institute for Theoretical Physics,University of Innsbruck, 6020 Innsbruck, Austria}
\author{Jan Chwede\'nczuk}
\affiliation{
  Faculty of Physics, University of Warsaw, ulica Pasteura 5, 02-093 Warszawa, Poland
}
\author{Marcin P{\l}odzie\'n}
\email{marcin.plodzien@qilimanjaro.tech}
\affiliation{
  Qilimanjaro Quantum Tech, Carrer de Veneçuela 74, 08019 Barcelona, Spain
}

\begin{abstract}
  Although entanglement can boost metrological precision beyond the standard quantum limit, the advantage often disappears with particle loss. 
  We demonstrate that scrambling safeguards precision by dispersing information about the encoded parameter into many-body correlations. 
  For Haar-random scrambling unitaries, we derive exact formulas for the average quantum Fisher information (QFI) of the reduced state after tracing out lost particles. 
  The result exhibits a threshold; any remaining subsystem larger than $N/2$ recovers the full QFI, while smaller subsystems contain negligible information. 
  We link this threshold to the scrambling-induced transition from area-law to volume-law entanglement and the associated growth of the Schmidt rank. 
  We outline two realizations—a brickwork circuit and chaotic XX-chain evolution—and demonstrate the protection of one-axis-twisted probes against the loss of up to half of the particles.
\end{abstract}

\maketitle

{\it Introduction.---}Quantum metrology uses nonclassical resources to precisely estimate parameters beyond 
the standard quantum limit~\cite{Giovannetti2006QuantumMetrology,Degen2017QuantumSensing,Pezze2018RMP}. 
Under ideal conditions, Greenberger-Horne-Zeilinger (GHZ) states can achieve Heisenberg scaling with respect to the number of particles $N$~\cite{BollingerEtAl1996,Pezze2009}. 
In practice, quantum advantage is fragile. Even a single lost particle can remove the coherence responsible for enhanced sensitivity. 
More broadly, local noise can constrain achievable scaling~\cite{HuelgaEtAl1997,Demkowicz2012}.

In this work we demonstrate that scrambling dynamics---i.e., state-rank growth under unitary evolution with support over an exponentially large set of 
Pauli strings~\cite{Haake2010, Hosur2016,KudlerFlam2022,Bhattacharyya2022,LoMonaco2024,Plodzien2025Haar}---can {\it protect} metrological information from the loss of up to $k < N/2$ particles.
We use the Hayden-Preskill protocol~\cite{HaydenPreskill2007}, in which scrambling dynamics enable the recovery of information from a subsystem after nearly half of its degrees of freedom are erased. 
We demonstrate the locking of the quantum Fisher information (QFI), which establishes the quantum Cram\'er–Rao bound~\cite{BraunsteinCaves1994,PezzeEtAl2018}, 
in the non-local quantum correlations of many-body systems. 
In this way, we establish a connection to quantum error correction (QEC). 
Since particle loss is an erasure error, random (scrambling) encoders are nearly optimal for the erasure channel~\cite{KnillLaflamme1997}.
QEC-inspired protocols have been developed for noise-robust metrology~\cite{LuYuOh2015,ArradEtAl2014,KesslerEtAl2014,ZhouEtAl2018NatComms,OuyangRengaswamy2023,Tan2019}. 
Random symmetric states display Heisenberg scaling that is robust to finite losses~\cite{OszmaniecEtAl2016PRX}.
We identify the mechanism behind the QFI-locking through the scrambling dynamics. It is due to an increase in the Schmidt rank and the associated change from an area law to a volume law in the quantum state.

We derive fully analytical expressions for the QFI of a subsystem after Haar-random scrambling. These expressions are valid for arbitrary initial states, whether they are simple product states or maximally entangled GHZ probes. 
We propose two mechanisms that allow for QFI locking 
in digital \cite{Brandao2016,Haferkamp2022} and analog \cite{Liang2025,Joshi2020} quantum simulators. These mechanisms involve applying scrambling unitaries to the initial information-encoded system.
Finally, we propose an experimental protocol that enables QFI locking in entangled states generated using the one-axis twisting (OAT) protocol.~\cite{KitagawaUeda1993, Wineland1994}. We demonstrate that the information encoded in the OAT states can be safeguarded against particle losses through scrambling dynamics produced by 
a chaotic XX spin chain with random transverse fields--—a system that can be easily implemented in existing quantum simulators.

Our findings offer a new perspective on many-body scrambling~\cite{Swingle2016,Swingle2018,Xu2024,LopezPiqueres2021,Burrell2009,Liang2025,Landsman2019,Blok2021,Google2025}
and chaos~\cite{DAlessio2016,Maldacena2016SYK,Maldacena2016bound,Deutsch1991,Srednicki1994}, which have recently been shown to be operationally useful for metrology~\cite{LiEtAl2023Science}. 
Experimental ``butterfly metrology'' demonstrations have shown that scrambling modifies local interactions, creating metrologically useful correlations.
Enhanced sensitivity is linked to Loschmidt echoes and out-of-time-order correlators~\cite{Ge2025,Hu2026}. 
Scrambling dynamics have also been suggested as a computationally efficient method for robust multiparameter estimation~\cite{Gong2026}. 
These approaches use scrambling to {\it generate} or {\it amplify} metrological sensitivity.

{\it Information encoding and locking.---}Consider a parameter $\theta$, encoded in a pure state $\ket{\psi(\theta)}\in\mathcal H$, followed by a global scrambling unitary $\hat U$. 
We derive the Haar-averaged QFI of the reduced density matrices that are obtained through the bipartitioning of the scrambled state using the Schmidt decomposition,
\begin{align}\label{eq.schmidt}
  \ket{\psi^{(U)}(\theta)}=\hat U\ket{\psi(\theta)} = \sum_{i=1}^{d_A}c_i\ket{\alpha_i}\otimes\ket{\beta_i}.
\end{align}
where $\ket{\alpha_i}\in\mathcal H_A$, $\ket{\beta_i}\in\mathcal H_B$, and $d_A\leqslant d_B$. The QFI quantifies the maximum amount of information that can be extracted about
$\theta$ via the Cram\'er--Rao bound, $(\Delta\theta)^2\geqslant\Id{\hat\varrho(\theta)}^{-1}$, and reads
\begin{align}\label{eq.qfi}
  \Id{\hat\varrho(\theta)}=2\sum_{i,j}\frac{\modsq{\bra{\psi_i}\dot{\hat\varrho}(\theta)\ket{\psi_j}}}{\lambda_i+\lambda_j},
\end{align}
where $\ket{\psi_i}$ are the eigenstates of $\hat\varrho(\theta)$ with the corresponding eigenvalues $\lambda_i$ and the summation runs through all $i$ and $j$ for which $\lambda_i+\lambda_j\neq0$.
The dot denotes the derivative of the density matrix with respect to $\theta$.

We proceed as follows: For both reduced density matrices $\hat\varrho_{A/B}(\theta)$, we calculate the corresponding $\Id{\hat\varrho^{(U)}_{A/B}(\theta)}$. Next, we average these two quantities
over the entire set of random matrices $\hat U$ using the Haar measure. The result is a pair of fully analytical expressions that can be compared to $\Id{\hat\varrho(\theta)}$, i.e., the amount of
information initially stored in $\ket{\psi(\theta)}$.

{\it Information spreading.---}Using Eq.~\eqref{eq.schmidt} we obtain
\begin{align}
  \hat\varrho^{(U)}_{A/B}(\theta)=\sum_{i=1}^{d_A}p_i\ket{\alpha_i/\beta_i}\bra{\alpha_i/\beta_i},
\end{align}
where $p_i=c_i^2$. According to Page's theorem~\cite{Page1993}, the typical reduced state of the smaller subsystem ($A$) is nearly maximally mixed \cite{Bengtsson2006}, $a_i\approx d^{-1}_A$, thus
\begin{align}\label{eq:page_approximation}
  \hat\varrho_A^{(U)}(\theta)\approx\frac{1}{d_A}\hat{\mathds1}_A.
\end{align}
In consequence, all non-zero eigenvalues of $\hat\varrho_B^{(U)}(\theta)$ are likewise equal to $d^{-1}_A$. However, since $d_A\leq d_B$, $\hat\varrho_B^{(U)}$ is, in general, 
not a full-rank operator in $\mathcal H_B$, i.e., 
\begin{align}
  \hat\varrho_B^{(U)}(\theta)\approx\frac{1}{d_A}\hat\Pi_B^{(U)},
\end{align}
where $\hat\Pi_B^{(U)}$ is a projector onto the subspace of $\mathcal H_B$ spanned by the eigenstates $\ket{\beta_i}$ with non-zero eigenvalues. 
Note that the precise make-up of this operator depends on $\hat U$.

\begin{figure}[t!]
    \centering
    \includegraphics[width=\linewidth]{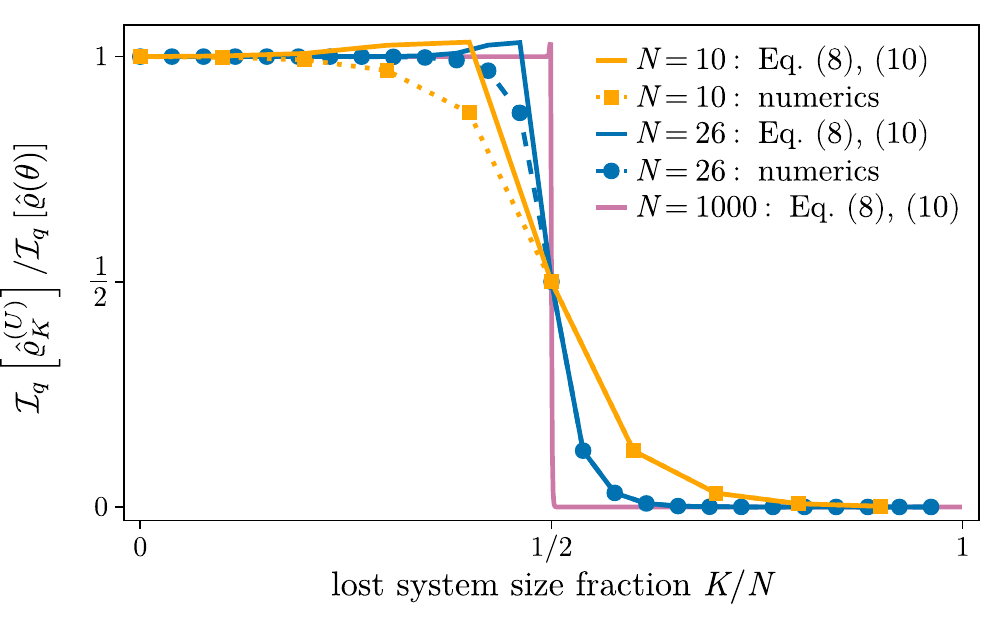}
    \caption{Typical value of the QFI for a reduced density matrix $\hat\varrho_K^{(U)}$ obtained after tracing-out $K$ out of $N$ qubits, normalized to the pure-state value, 
      $\Id{\hat\varrho_K^{(U)}}/\Id{\hat\varrho(\theta)}$. 
      Solid lines correspond to the analytical prediction in Eq.~\eqref{eq:qfi_a_specific} and Eq.~\eqref{eq:qfi_b_specific} for $N=10$ and 26, The  squares andcircles represent numerical results. 
      For comparison, we display the analytical curve for $N=1000$ qubits.}
    \label{fig:haar}
\end{figure}

Plugging the approximation~\eqref{eq:page_approximation} into Eq.~\eqref{eq.qfi} we obtain
\begin{align}\label{label.qfi.A}
  \Id{\hat\varrho^{(U)}_{A}}\approx d_A\tr_A\left[\left(\dot{\hat\varrho}_A^{(U)}\right)^2\right],
\end{align}
where, to shorten the notation, we omitted the dependence of the density matrix on $\theta$.
Next, we utilize the SWAP-\textit{trick}~\cite{Werner1989, Burhman2001}, which for $\hat X$ and $\hat Y$ acting on the same $\mathcal H_0$ is $\tr[\hat X\hat Y]=\tr[(\hat X\otimes \hat Y)\hat S]$,
where the trace on the right-hand side is taken over $\mathcal H_0^{\otimes 2}$. The SWAP operator acts on $\mathcal H_0^{\otimes 2}$ as well:  $\hat S(\ket{\psi}\otimes\ket{\phi})=\ket{\phi}\otimes\ket{\psi}$. 
Introducing $\hat{\tilde{\mathds 1}}_B=\hat{\mathds 1}_B\otimes\hat{\mathds 1}_B$, the QFI from Eq.~\eqref{label.qfi.A} written in terms of the trace over $\mathcal H^{\otimes2}$ becomes
\begin{align}\label{eq:qfi_a_twocopy}
  \Id{\hat\varrho^{(U)}_{A}}\approx d_A\tr\left[\left(\dot{\hat\varrho}_A^{(U)\otimes2}\right)(\hat S_A\otimes\hat{\tilde{\mathds 1}}_B)\right].
\end{align}
We average this result over the complete set of unitary matrices by integrating with the Haar measure $\mu(U)$. Using the Weingarten calculus~\cite{Weingarten1978,Collins2006,Mele2024,endm}, 
the outcome is a surprisingly compact and illuminating expression
\begin{align}\label{eq:qfi_a_specific}
  \hvg{\Id{\hat\varrho^{(U)}_{A}}}=\Id{\hat\varrho(\theta)}\times f_A,
\end{align}
where the rational function $f_A=f_A(d_A,d_B)$ satisfies
\begin{align}\label{eq:qfi_a_expanded}
  f_A=\frac{d(d_A^2-1)}{2(d^2-1)}\ \ \underset{d_A,d_B\gg1}{\approx}\ \ \frac{d_A}{2d_B},
\end{align}
The last step assumes $d_A,d_B\gg1$. Clearly, if $d_B$ dominates over $d_A$ (which happens even if it consists of just a few qubits more), the amount of information
that can be extracted from $A$ is very small. Only if $d_A\sim d_B$, the loss is tolerable.

The main result of this work is that the complementary subsystem $B$ retains the full QFI. Applying analogous techniques to $\hat\varrho^{(U)}_{B}$, we obtain~\cite{endm}
\begin{align}\label{eq:qfi_b_specific}
  \hvg{\Id{\hat\varrho^{(U)}_{B}}}&=\Id{\hat\varrho(\theta)}\times f_B
\end{align}
where the function $f_B=f_B(d_A,d_B)$ is
\begin{align}\label{eq:qfi_b_expanded}
  f_B&=\frac{d_A^3(d_B^2-1)\left[d_B(2d+9)+d_A(-3d_A^2+d+15)\right]}{2(d^2-1)(d+2)(d+3)}\nonumber\\
  &\underset{d_B\gg d_A\gg1}{\approx}1.
\end{align}
Equations~\eqref{eq:qfi_a_specific} and \eqref{eq:qfi_b_specific} constitute the central analytical result: scrambling locks the encoded information into volume-law correlations such that any subsystem retaining more than half the particles ($d_B > d_A$) recovers the full QFI. This holds for arbitrary initial states $\ket{\psi(\theta)}$.

\begin{figure}[t!]
    \centering
    \includegraphics[width=\linewidth]{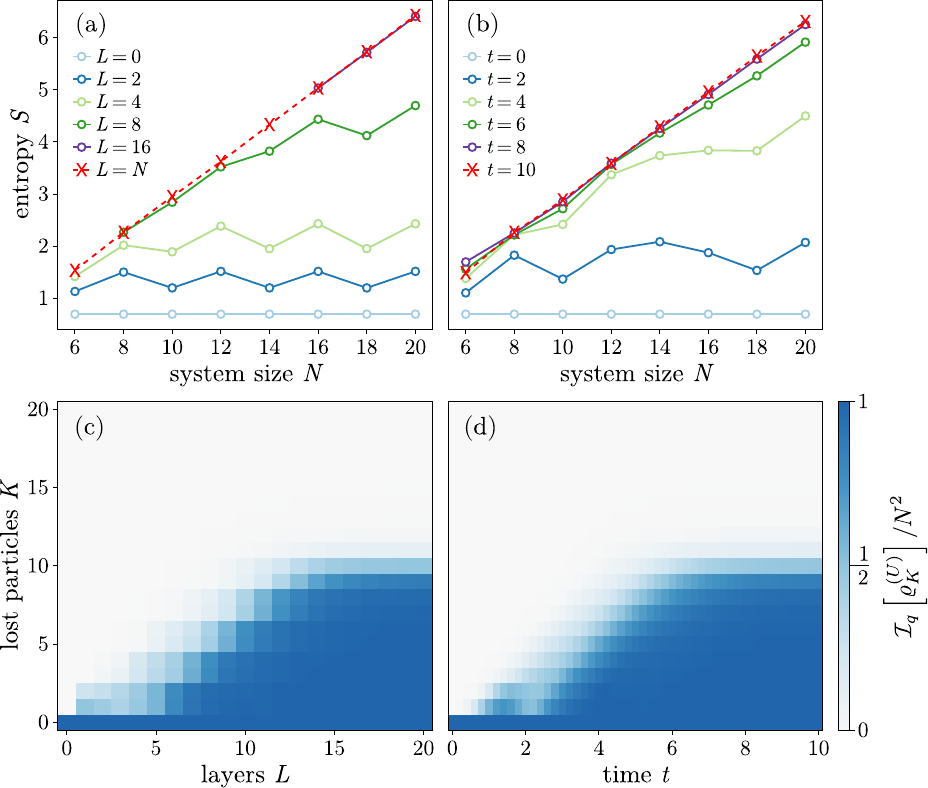}
    \caption{Panels (a)-(b): Bipartite entanglement entropy $S$ at the central bipartition as a function of system size $N$ for various circuit depths $L$ (panel (a)), 
      and analog evolution time $t$ (panel (b)). Circuits with $L\le N/2$ have higher Schmidt rank, converging to the maximal value $\chi_{\rm max}=2^{N/2}$ (dashed line) for $L=N$. Panels (c) - (d): 
      The QFI after tracing-out $K$ qubits for increasing circutis depths $L$, panel (c), and analog evolutions timse $t$, panel (d). 
      The dark regions indicate preserved Heisenberg-limited QFI. 
      As $L$ ($t$) increases, QFI protection extends to larger $k$, demonstrating that volume-law entanglement locks the metrological information against particle loss.}
    \label{fig:area_to_volume_law}
\end{figure}

We verify our findings numerically by simulating a system of $N$ qubits, initialized in a product state $\ket{0}^{\otimes N}$, 
and the parameter $\theta$ is encoded via the generator $\hat{G}=\frac12\sum_i\hat\sigma_i^x$, yielding the initial pure-state QFI value of $\Id{\hat\varrho(\theta)}=N$. Following this encoding, the state is scrambled by a Haar-random unitary $\hat U$, and we compute the QFI of the reduced density matrix $\hat\varrho_K^{(U)}$ obtained by tracing out $K\in\{0,1,\ldots,N\}$ qubits. The results are shown in Fig.~\ref{fig:haar}, where solid lines represent our analytical prediction expressed by Eq.~\eqref{eq:qfi_a_specific} and Eq.~\eqref{eq:qfi_b_specific}, and squares and circles correspond to the numerically sampled $\hat U$. It is important to note that while the analytical curves represent an ensemble average with the Haar measure, the numerical results for individual realizations converge tightly to these values. This is a hallmark of measure concentration and typicality, which ensure that the properties of a single, randomly sampled high-dimensional state are highly representative of the entire ensemble average.

The QFI locking mechanism is rooted in the entanglement structure. A GHZ state has Schmidt rank $\chi=2$ across every bipartition, and its encoded information resides in a fragile low-rank coherence,
that can be easily destroyed by tracing out even a single particle. The scrambling process transforms the fragile rank-2 structure into a state with nearly maximal Schmidt rank ($\chi \sim 2^{N/2}$), 
spreading the encoded information uniformly across all degrees of freedom. 


{\it Scrambling unitaries.---}Although the Haar-random unitaries provide an analytically tractable model, they are difficult to realize experimentally. 
Nevertheless, the underlying mechanism---growth of the Schmidt rank---can be achieved using protocols that are accessible to experimentation.
We consider two unitaries: the first is a quantum circuit of $L$ layers $\hat{U}(L) = \prod_{\ell=1}^L\hat U^{(\ell)}$, where
\begin{align}
  \hat U^{(\ell)}=\prod_{i=1}^N \hat{R}_i^{(y)} \prod_{i\in\mathrm{e}}\hat{\mathrm{CX}}_{i,i+1}\prod_{i\in\mathrm{o}}\hat{\mathrm{CX}}_{i,i+1}
\end{align}
consisting of  rotational gates $\hat{R}^{(y)}$ by $\pi/4$, followed by brickwork-topology of two-qubit entangling control-X ($\hat{\rm CX}$) gates (e/o is a set of even/odd indeces).

The second one is generated by the non-integrable chaotic  XX model with transverse random field \cite{Muller1987} 
\begin{equation}\label{eq:xx_hamiltonian}
    \hat H = \sum_{i=1}^{N-1}(\hat{\sigma}^x_i\hat{\sigma}^x_{i+1}+\hat{\sigma}^y_i\hat{\sigma}^y_{i+1})+\sum_{i=1}^Nh_i\hat{\sigma}^x_i,
\end{equation}
where $h_i$ are drawn with a flat distibution over $[-1, 1]$, independently for each spin.
Both unitaries transform a low-rank, area-law entangled state into a scrambled, high-rank, volume-law state as the number of layers ($L$) or evolution time ($t$) increases. 
This provides resources for QFI locking in non-local correlations.

To gain insight into QFI-locking scrambling mechanisms, we will consider the following protocol. We encode the information $\theta$ using the generator $\hat G = \frac12\sum_i\hat{\sigma}^z_i$ 
in an initial GHZ state of $N$ qubits, and then apply a unitary $\hat{U}(L)$ (for digital protocol) or unitary $\hat{U}(t) = e^{-i t\hat{H}}$ (for analog protocol). 
We demonstrate that the QFI remains protected against qubit loss when unitaries generate states that follow volume-law entanglement scaling.

The QFI locking mechanism is shown in Fig.~\ref{fig:area_to_volume_law}. 
Panels (a) and (b) show how the mid-cut bipartite entanglement entropy scales with system size for an increasing number of layers, $L$, in panel (a), 
and for an increasing analog evolution time, $t$, in panel (b). 
Shallow circuits ($L\ll N$) and short times result in area-law states. For deeper circuits ($L\ge N/2$) and longer evolution times, the state becomes highly scrambled, 
following the volume law of bipartite entanglement entropy scaling.
Through the scrambling unitaries, the classical information encoded in a low-rank GHZ state becomes encoded in a high-rank state among many non-local correlations. This protects the QFI against particle loss. 
The bottom panels show the QFI phase diagram in the circuit depth-particle loss plane, panel (c), and the scrambling time-particle loss plane, panel (d), for 20 qubits.
At $L=0$ ($t = 0$) any particle loss destroys the QFI encoded in the rank-2 GHZ state. 
As $L$ increases, the protected region expands.
For circuits that are deep enough ($L > N/2$), the full QFI is preserved for all $k < N/2$. This is in agreement with the prediction of the Haar measure.
A similar transition occurs in analog evolution for $t>5$. The transition from area-law entanglement ($\chi=2$ for GHZ) to volume-law entanglement directly controls QFI protection.

{\it Illustration: OAT systems.---}We will now demonstrate how these results apply to an important family of states generated by OAT dynamics, i.e.,
\begin{align}
  \ket{\psi_\tau}=e^{-i\tau\hat J_z^2}\ket0_x^{\otimes N},
\end{align}
where $\tau=\chi t$ is the interaction time in units of the two-body coupling strength $\chi$ and $\hat J_z=\frac{1}{2}\sum_{i=1}^N\hat \sigma^z_i$. 
The OAT dynamics generates metrologically useful many-body entangled, and many-body Bell correlated states~\cite{Plodzien2022}, as well as nonstabilizer states \cite{Hernande2025OATSRE}. 
Their robustness  depends strongly on the evolution time: for $\tau \lesssim N^{-2/3}$, 
the protocol produces spin-squeezed states that offer enhanced sensitivity while remaining comparatively resilient to finite detector resolution, moderate particle loss, and dephasing~\cite{Pezz2018}. 

At longer times, oversqueezed non-Gaussian states emerge with increased entanglement depth. These states are generally robust against 
detection noise but they become significantly more vulnerable to dephasing~\cite{Davis2016,Nolan2017,Baamara2021}. 

For $\tau \simeq \pi/n$ (with even $n$), the system approaches a macroscopic superposition of $n$ maximally separated spin-coherent components, i.e., a multi-component cat (``kitten'') state. 
This state is relatively tolerant to loss and can interpolate between sub-shot-noise and near-Heisenberg sensitivities~\cite{Tatsuta2019}, 
albeit with pronounced fragility to phase noise~\cite{Ferrini2008,Spehner2014,Huang2015,Chalopin2018}.
Finally, near $\tau \simeq \pi/4$, the evolution yields an $x$-oriented GHZ state that achieves Heisenberg scaling but is highly susceptible to noise.

\begin{figure}[t!]
    \centering
    \includegraphics[width=\linewidth]{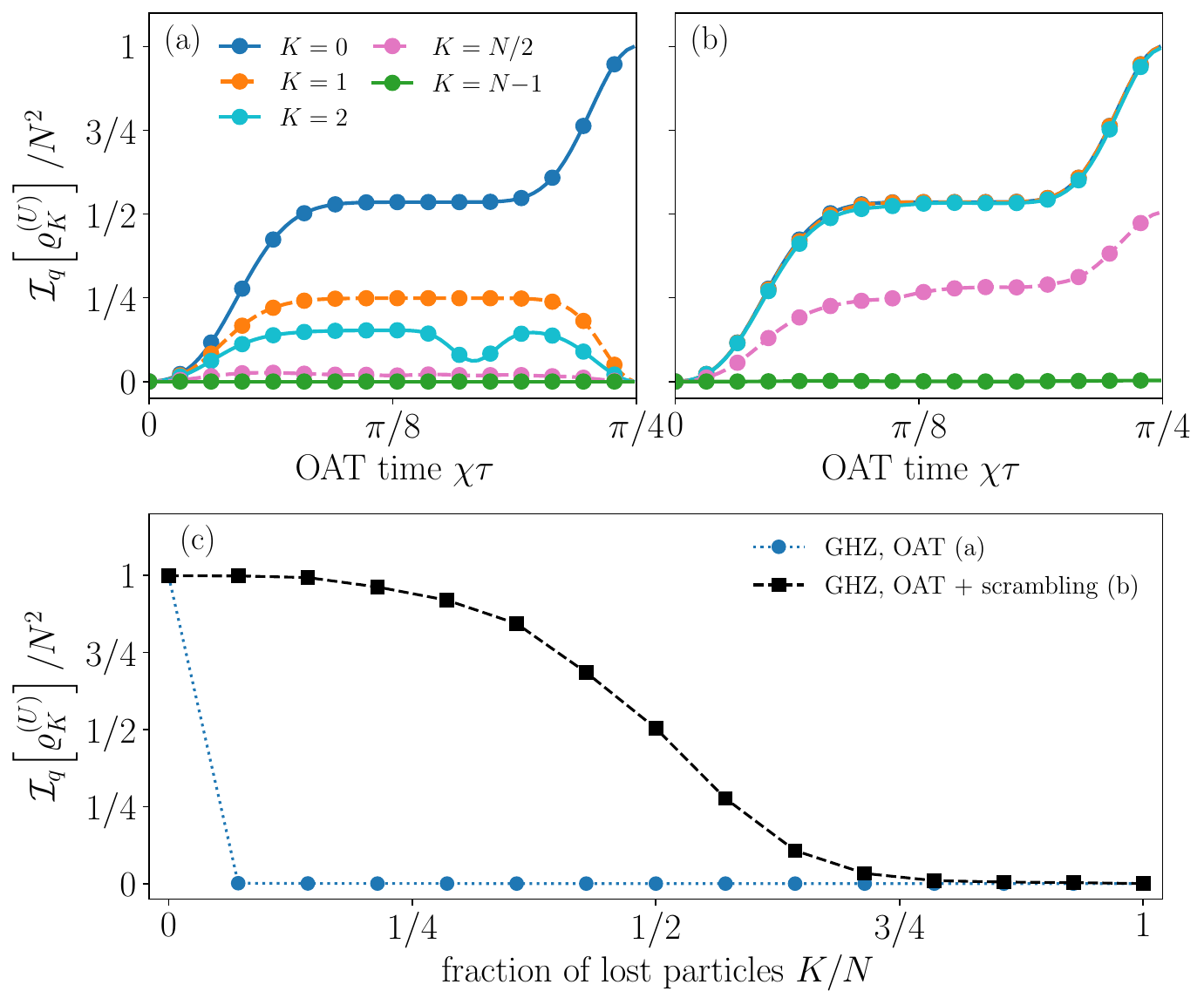}
    \caption{(a): the QFI for OAT-generated states as a function of time after tracing out $K$ qubits. Blue solid line correspond to pure OAT state without particle loss ($K=0$). 
      The lines (from top to bottom) corresponds to QFI in a reduced state after tracing out $K = 1, 2, N/2, N/2-1$ qubits ($N=20$), indicating diminishing metrological quantum advantage. 
      (b): the QFI for OAT-generated states followed by analog scrambling evolution, resulting in QFI-locking. The QFI is protected against particle loss for $K<N/2$, as illustrated
      by (c), which plots the QFI ath $\tau=\pi/4$ as a function of the fraction of lost partiles. }
    \label{fig:oat}
\end{figure}

For OAT-generated state at each time $\tau$ the information $\theta$ is encoded through the generator $\hat G = \frac{1}{2}\sum_i \hat\sigma^x_i$
\begin{align}
  \ket{\psi_\tau(\theta)}=e^{-i\theta\hat G}\ket{\psi_\tau}.
\end{align}
The QFI calculated using Eq.~\eqref{eq.qfi} fed with the above state is shown by the top line in Fig.~\ref{fig:oat}(a) for $\tau\in[0,\pi/4]$ and $N=20$. 
At time $\tau=\pi/4$, the GHZ state gives the Heisenberg scaling, $\mathcal I_q=N^2$. This state is not reliable for metrology, 
since the loss of just a single qubit erases all information about $\theta$ and causes the reduced-state QFI to vanish completely. 
This illustrated by the remaining curves in Fig.~\ref{fig:oat}(a), which vary according to the number of lost qubits.

The QFI locking protocol employs the unitary evolution,  $\ket{\tilde{\psi}_\tau(\theta)} = \hat U (t_{\rm scr})\ket{\psi_\tau(\theta)}$,
generated by the chaotic XX Hamiltonian~\eqref{eq:xx_hamiltonian} with $t_{\rm scr} = 20$.
Fig \ref{fig:oat}(b) presents QFI particles loss protected states $|\tilde{\psi}_\tau(\theta)\rangle$. As in Fig.~\ref{fig:oat}(a), the color of each line corresponds to the number of lost particles.
Indeed, the QFI encoded in the GHZ state generated by the OAT and then subjected to scrambling dynamics is protected against particle losses for $K < N/2$. 
This is further illustrated in Fig.~\ref{fig:oat}(c), which displays the loss of QFI for the GHZ state as a function of the ratio of lost particles, $K/N$.

{\it Conclusions.---}We have demonstrated that scrambling dynamics are a robust mechanism that protects quantum-enhanced metrological information against particle loss. 
Our central analytical result is that, for any initial state, Haar-random scrambling preserves quantum Fisher information in any subsystem comprising more than half of the particles. 
This effect occurs due to the transition from area-law to volume-law entanglement, which disperses information about $\theta$ across all degrees of freedom.
We proposed two experimentally feasible protocols that demonstrate how practical scramblers can achieve near-Haar performance while retaining full metrological advantage 
for losses of up to half of the total particles. As a concrete example, we demonstrated how the protocol functions with metrologically useful states generated via the one-axis twisting method. 
Furthermore, our results demonstrate that scrambling dynamics can protect quantum-enhanced sensitivity against erasure errors. 
This work paves the way for fault-tolerant quantum sensing in platforms where particle loss is the primary decoherence mechanism. 
These platforms include atomic ensembles, photonic networks, and noisy intermediate-scale quantum devices.

{\it Acknowledgements.---}This work was supported by the National Science Centre, Poland, within the QuantERA II Programme that has received funding from the European Union’s Horizon 2020 
research and innovation programme under Grant Agreement No 101017733, Project No. 2021/03/Y/ST2/00195.
 
The data that support all the figures are openly available~\cite{data}. The code to generate the data reported in this work is available at~\cite{data2}.

\section{END MATTER}

\setcounter{equation}{0}                                                                                                
\renewcommand{\theequation}{EM.\arabic{equation}}                                                                        

Below, we outline the analytical calculation that yields the expressions given by Eq.~\eqref{eq:qfi_a_expanded} and Eq.~\eqref{eq:qfi_b_expanded} of the main text. 
First, we calculate the QFI for the smaller subsystem, where the calculation is simpler. Then, we introduce all the techniques used to obtain the final result. 
Then, we generalize these techniques to perform the calculation for the larger subsystem, highlighting the subtle differences between the two cases that must be accounted for.

\subsection{QFI for the smaller subsystem}

We start from Eq.~\eqref{eq:qfi_a_twocopy}, which states that, under Page's approximation, the QFI for subsystem A is given by
\begin{align}\label{eq:qfi_a_twocopy_endmatter}
  \Id{\hat\varrho^{(U)}_{A}}\approx d_A\tr\left[\left(\dot{\hat\varrho}_A^{(U)} \otimes \dot{\hat\varrho}_A^{(U)} \right)(\hat S_A\otimes\hat{\tilde{\mathds 1}}_B)\right].
\end{align}
For notational convenience, we introduce the operators
  \begin{align}
    \hat C=\dot{\hat\varrho}(\theta),\ \ \hat C^{(U)}=\hat U \hat C \hat U^\dagger,\ \ \hat C^{(U)}_{m}=\dot{\hat\varrho}_{m}^{(U)},
\end{align}
with $m=A/B$. By linearity, the average of \eqref{eq:qfi_a_twocopy_endmatter} is 
\begin{align}\label{eq:hvg_qfi_a}
  &\hvg{\Id{\hat\varrho^{(U)}_{A}}}=d_A\tr\left[\hvg{\hat C^{(U)\otimes 2}_A}(\hat S_A\otimes\hat{\tilde{\mathds 1}}_B)\right]\nonumber\\
  &=d_A\tr\left[\hvg{\hat U^{\otimes 2}(\hat C^{\otimes 2})\hat U^{\dagger\otimes 2}}(\hat S_A\otimes\hat{\tilde{\mathds 1}}_B)\right].
\end{align}
In order to evaluate the expectation value we use the Weingarten calculus toolkit \cite{Mele2024}. The $n$-th moment operator $\hat M_n(\hat O)$, defined as
\begin{equation}
    \hat M_n(\hat O) = \hvg{\hat U^{\otimes n}\hat O\hat U^{\dag\otimes n}},
\end{equation}
is given by
\begin{equation}
    \hat M_n(\hat O) = \sum_{\sigma,\tau\in S_n}\mathrm{Wg}(\tau\sigma)\tr[\hat O\hat V(\tau)]\hat V(\sigma),
\end{equation}
where $\sigma,\tau$ are elements of the symmetric group $S_n$, i.e., permutations of $n$ elements. The $\hat V(\sigma)$ is the unitary permutation operator corresponding to a permutation $\sigma$, 
\begin{align}
    \hat V(\sigma)\bigotimes_{k=1}^n\ket{\psi_k}\equiv\bigotimes_{k=1}^n\ket{\psi_{\sigma(k)}},
\end{align}
Moreover, $\hat O$ is an operator acting on $n$ copies of a Hilbert space of dimension $d$, and $\mathrm{Wg}(\sigma)$ is the Weingarten function of a permutation $\sigma$, which for a fixed $\sigma$ is a rational function that can be expressed solely via the number $d$.

For $n=2$, we have $S_2=\{e,(12)\}$, where the cycle notation $(12)$ denotes the transposition $1\leftrightarrow2$. The corresponding permutation operators are the identity $\mathbb{I}$ and 
the SWAP operator $\hat S$, each acting on two copies of the entire AB system.

Due to the special structure of the operator $\hat C$, we have $\tr[\hat C] = 0$ and $\tr[\hat C^2] = \Id{\hat\varrho(\theta)}/2$. Using this fact and substituting the explicit forms of the Weingarten functions, we get
\begin{equation}
    \hvg{\hat C^{(U)\otimes2}} = \frac{\Id{\hat\varrho(\theta)}}{2}\left[\frac{1}{d^2-1}\hat S-\frac{1}{d(d^2-1)}\hat{\tilde{\mathds 1}}\right].
\end{equation}
By plugging this expression into Eq.~\eqref{eq:hvg_qfi_a}, evaluating the traces, and using $d=d_Ad_B$, we obtain
\begin{equation}
    \hvg{\Id{\hat\varrho^{(U)}_{A}}} = \Id{\hat\varrho(\theta)}\frac{d(d_A^2-1)}{2(d^2-1)},
\end{equation}
as reported in the main text.

\subsection{QFI for the larger subsystem}

We now switch to the larger subsystem and denote by $\lambda_i$ the eigenvalues of $\hat\varrho_B^{(U)}$ and by $\ket{\psi_i}$ the corresponding eigenstates. 
Note that the general expression for $\Id{\hat\varrho^{(U)}_{B}}$ is
\begin{equation}
    \Id{\hat\varrho^{(U)}_{B}} = 2\sum_{i,j}\frac{\modsq{\bra{\psi_i}\dot{\hat\varrho}^{(U)}_B\ket{\psi_j}}}{\lambda_i+\lambda_j},
\end{equation}
where the sum runs over pairs of eigenvalues $(\lambda_i,\lambda_j)$ such that $\lambda_i+\lambda_j\neq 0$. The crucial step is to observe that this sum can be decomposed into three parts characterized by $(\lambda_i\neq 0, \lambda_j\neq 0)$, $(\lambda_i\neq0, \lambda_j=0)$, and $(\lambda_i=0,\lambda_j\neq 0)$. Using the fact that
\begin{align}
    \sum_{i:\lambda_i\neq0}\ket{\psi_i}\bra{\psi_i}&\equiv\hat\Pi_B^{(U)} \\
    \sum_{i:\lambda_i=0}\ket{\psi_i}\bra{\psi_i} &= \hat{\mathds 1}_B-\hat\Pi_B^{(U)},
\end{align}
we may rewrite $\Id{\hat\varrho^{(U)}_{B}}$ in terms of traces over $\mathcal H_B$ with appropriately placed projectors as
\begin{align}
    \Id{\hat\varrho^{(U)}_{B}} &= 4d_A\tr\left[\hat C_B^{(U)}\hat C_B^{(U)}\hat\Pi_B^{(U)}\right] \nonumber\\
    &-3d_A\tr\left[\hat C_B^{(U)}\hat\Pi_B^{(U)}\hat C_B^{(U)}\hat\Pi_B^{(U)}\right].
\end{align}
To proceed, we use the generalized version of the SWAP-trick, namely
\begin{equation}\label{eq:general_swap_trick}
  \tr\left[\prod_{i=1}^n\hat X_i\right] = \tr\left[\hat V\Big((12\ldots n)\Big)\bigotimes_{i=1}^n\hat X_i\right],
\end{equation}
where $(12\ldots n)$ is the full $n$-cycle permutation, $1\rightarrow2\rightarrow\ldots\rightarrow n\rightarrow 1$.
\begin{widetext}
Substituting $\hat\Pi_B^{(U)}=d_A\hat\varrho_B^{(U)}$ and using Eq.~\eqref{eq:general_swap_trick}, we get
\begin{equation}
    \Id{\hat\varrho^{(U)}_{B}} = 4d_A^2\tr\left\{\left[\hat C^{(U)\otimes2}\otimes \hat\varrho^{(U)}\right]\left[\hat{\mathds1}_A\otimes \hat V_B((123))\right]\right\}
    -3d_A^3\tr\left\{\Big(\hat C^{(U)}\otimes\hat\varrho^{(U)}\Big)^{\otimes2}\left[\hat{\mathds1}_A\otimes \hat V_B((1234))\right]\right\}
\end{equation}
To evaluate the Haar average $\hvg{\Id{\hat\varrho^{(U)}_{B}}}$, we once again use the general $n$-th moment Weingarten formula for $n=3$ and $n=4$. After counting the permutations that add non-vanishing contributions and computing the relevant three-copy and four-copy traces, we eventually find
\begin{align}
    \hvg{\Id{\hat\varrho^{(U)}_{B}}} &= \Biggl\{\sum_{\sigma\in S_3}d_A^{\,c(\sigma)+2}d_B^{\,c(\sigma(123))}\Bigl[2\mathrm{Wg}((12)\sigma)+\mathrm{Wg}((123)\sigma)+\mathrm{Wg}((132)\sigma)\Bigr] \nonumber\\
    &-\sum_{\sigma\in S_4}d_A^{\,c(\sigma)+3}d_B^{\,c(\sigma(1234))}\Biggl[\frac32\sum_{\tau\in S_4'}\mathrm{Wg}(\tau\sigma)+\frac34\sum_{\tau\in S_4''}\mathrm{Wg}(\tau\sigma)\Biggr]\Biggr\},
\end{align}
where $S_4'=\{(13),(13)(24)\}$, $S_4''=\{(123), (132), (134), (143), (1243), (1324), (1342), (1423)\}$, and $c(\sigma)$ is the number of cycles in the unique disjoint-cycle decomposition of a permutation $\sigma$.

The common property of all permutations that give non-vanishing contributions is that they do not leave any of the two $\hat C$ operators unpermuted, which would lead to the multiplication 
by $\tr(\hat C)=0$. After substituting the explicit forms of the Weingarten functions and performing summations over permutations, we find
\begin{equation}
    \hvg{\Id{\hat\varrho^{(U)}_{B}}} = \Id{\hat\varrho(\theta)}\frac{d_A^3(d_B^2-1)\left[d_B(2d+9)+d_A(-3d_A^2+d+15)\right]}{2(d^2-1)(d+2)(d+3)},
\end{equation}
as reported in the main text.
\end{widetext}

\end{document}